\documentclass{article}
\usepackage{spconf,amsmath}
\usepackage{amssymb}
\usepackage{amsthm}
\newtheorem{theorem}{Theorem}
\usepackage[pdftex]{graphicx}
\usepackage{epstopdf}
\usepackage[caption=false,font=footnotesize]{subfig}

\def\L{{\cal L}}

\title{Minimax Game-Theoretic Approach to\\Multiscale H-infinity Optimal Filtering}
\name{Hamza Anwar and Quanyan Zhu\thanks{This work is partially supported by the grants EFMA-1441140 and SES-1541164 from National Science Foundation.}}
\address{Electrical and Computer Engineering Department, New York University, Brooklyn, NY 11201, USA\\{Email: \{ha1082, qz494\}@nyu.edu}
}
\begin{document}
\ninept
\maketitle
\begin{abstract}
Sensing in complex systems requires large-scale information exchange and on-the-go communications over heterogeneous networks and integrated processing platforms. Many networked cyber-physical systems exhibit hierarchical infrastructures of information flows, which naturally leads to a multi-level tree-like information structure in which each level corresponds to a particular scale of representation. This work focuses on the multiscale fusion of data collected at multiple levels of the system. We propose a multiscale state-space model to represent multi-resolution data over the hierarchical information system and formulate a multi-stage dynamic zero-sum game to design a multi-scale $H_{\infty}$ robust filter.  We present numerical experiments for one and two-dimensional signals and provide a comparative analysis of the minimax filter with the standard Kalman filter to show the improvement in signal-to-noise ratio (SNR).
\end{abstract}
\vspace*{-2mm}
\section{Introduction}
\label{sec:intro}
Data processing of large-scale complex systems requires massive information exchange and communications that are often organized in a hierarchical manner in which data collected at multiple scales represent different resolutions. It is imperative that the fusion of the sensed data need to be cost-effective, energy-efficient, quality-ensuring, vulnerability-evasive, and overall robust in the process. Hence in this work, we aim to design a multi-scale minimax filter that allows state estimation which is robust to disturbances across multiple scales of the data collection.

To this end, we first build a multiscale state-space representation of the multiscale data collection process. Leveraging the state-space framework for modeling stochastic phenomenon at multiple scales in Chou et. al. \cite{chou1991stochastic}, we formulate the robust multiscale state estimation problem using a zero-sum dynamic game approach. The development of multiscale state estimation algorithms provides a spatially dynamic approach towards fusing the multi-layer data and enables a scalable and implementable solution for the large-scale system.

To illustrate the multiscale model, we can consider crime incident reporting as an example. Crime incidents are reported at nearest precincts in the city. All precincts in a city have the aggregated criminal activity information of the whole city. Cities within a state can report this information at the state level. Similarly, the states of the country can aggregate criminal activity information from cities and report it to the federal government. This architecture can be considered as a pyramidal data structure with all precincts' data at the finest level, city data at one coarser level, state data at one more coarser level, and so on. The data at each successive level is an aggregated representation of the same data, and thus data at each level is linearly related. Moreover, when the available data is noisy, a scheme for filtering out the unnecessary cofactors is often required. Optimal filtering minimizes the error between estimated and actual values across all levels while making the estimator robust to noise. Data is fused along coarser levels and interpolated along finer levels. This work formulates a filtering problem of this nature, where data are collected at multiple layers of different resolutions and the granularity of the information increases as the system zooms into finer levels.

\vspace*{0mm}
We present a robust $H_\infty$ filter that outperforms regular Kalman filter when the multiscale system is subject to additive noise in input and measurements. The filter operates in the direction of coarse-to-fine dynamics. Moreover, we extend our $H_\infty$ filter to incorporate measurements at $k$-th level when estimating the state at $k$-th level (in contrast to a predictor-based design that relies on one previous step measurements). The use of game-theoretic $H_\infty$ filter is not new in signal processing schemes \cite{shen1997game}; however, up till now, it has not found applications in multi-resolution information systems because the tool of multiscale system modeling has not been widely used in all relevant domains owing to its complex nature \cite{ferreira2007multiscale}. In our work, we show the performance benefits of using the minimax robust $H_\infty$ filter in contrast to Kalman filter for 1-D and 2-D signals.

Multiscale modeling has been well studied in the statistics and signal processing communities \cite{clippingdale1989least,willsky2002multiresolution}, e.g., the development of wavelet analysis \cite{ferreira2007multiscale} and its applications in image compression and image restoration \cite{banham1996spatially}. Among the relevant uses of multiscale models for estimation tasks are in geoscience and remote sensing especially when there are heterogeneous suites of sensors (infrared, visual, microwave, etc.) \cite{zheng2008remote} and in hierarchical graphical modeling in machine learning \cite{ahmed2013hierarchical}.
In recent works for multi-layered systems, {\em transcale} optimal controller and robust state-estimator for discrete-time systems has been proposed by Zhao et. al. \cite{zhao2015transcale,zhao2013robust}. Their theory uses wavelets decomposition to relate signals across resolution levels. The focus of our work, being starkly different, is in the fusion of sensor data of different resolutions as in \cite{zhang2004multiresolution}. Our work specifically deals with the multiscale data fusion problem that arises from the spatial hierarchical architectures of the sensors  in contrast to time-domain multiscale filtering problems.
\vspace*{-2mm}
\section{Multiscale State-Space Estimator Design}
\label{sec:multiscale}
\begin{figure}[t]
\centering
\includegraphics[width=\linewidth]{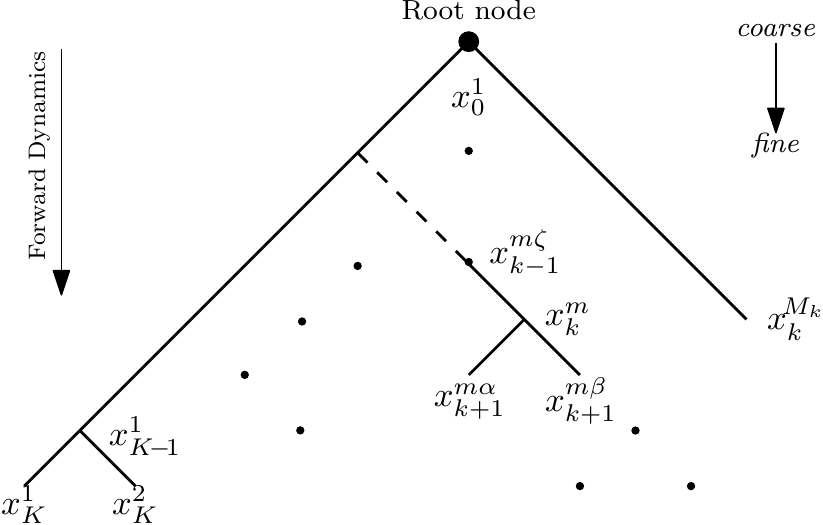}
\caption{Depiction of the dyadic tree structure for a multiscale linear system}\label{fig:tree}
\end{figure}
Consider a dyadic tree structure of states: nodes are arranged in a
layered setting. Fig.~\ref{fig:tree} represents $K+1$ levels of a latent process on a dyadic tree. At the $k$-th level the (vectorized) process is denoted by $x_k$ and its corresponding state-vector at $m$-th node is $x_k^m \in \mathbb{R}^n$. Thus, the multiple layers of the tree indicate different representations of the complete latent process $x$. We assume that the nodes of the latent process at a given level are conditionally independent of each other given the state at an immediate coarser level. Our tree structure has the following form of the forward dynamics (coarse-to-fine):
\begin{align}
\label{eq:model}
\begin{split}
x_{k+1}^{m\alpha} &= A_{k+1}^{m\alpha} x_{k}^{m} + B_{k+1}^{m\alpha} w_{k+1}^{m\alpha},\\
x_{k+1}^{m\beta} &= A_{k+1}^{m\beta} x_{k}^{m} + B_{k+1}^{m\beta} w_{k+1}^{m\beta},\\
y_k^m &= C_k^m x_k^m + v_k^m,\\
\end{split}
\end{align}
\[k\in \mathcal{K}:=\{0,\dots,K-1\};~m\in\mathcal{M}_k:=\{1,\dots,M_k\};~x_0^1 = x_0,\]
where $v_k^m \in \mathbb{R}^p$ is the measurement noise with covariance given as $\mathbb{E}[v_k^mv_k^{m\top}]= R_k^m$, and $w_k^m \in \mathbb{R}^q$ the disturbance at $m(k)$-th node, having $\mathbb{E}[w_k^mw_k^{m\top}]= I$ w.l.o.g. Here, $A_k^m$ represents the interpolation matrix, i.e., the relationship between system states at coarse and fine levels. $B_k^m$ is the input matrix for disturbance signals. Along coarse-to-fine recursion, $w$ signals actually represent the higher details added, but because coarse-to-fine recursion is analogous to the multiresolution synthesis of signals, $w$ is rather interpreted as the disturbance signal corrupting our state trajectories $x$ along resolution scales. For dyadic tree, the number of nodes at $k$-th level is  $M_k = 2^k$. Since the coarsest level only has a single (root) node, we represent state of root node $x_0^1$ as merely $x_0$. For each $m$, $m\alpha:=2m-1$ and $m\beta:=2m$ indicate its two children nodes in the next finer level and $m\zeta:=\lceil{\frac{m}{2}}\rceil$ is its parent node at a coarser level. We assume that $\{w^m\}$, i.e., the disturbance in $m$-th path along coarse-to-fine scales, is any $l_2$ sequence\footnote{The set representing the path from $w_m^K$ to the root is $\{w^m\} = \{w^m_K, w_{K-1}^{m\zeta},w_{K-2}^{(m\zeta)\zeta},\cdots,w_0^1\}$. The sum of $l_2$ norms of each $w_k^m$ along the path that leads to $w_K^m$ from the root is bounded.}.

Ultimately, instead of estimating system states $x_k^m$, we are interested in estimating a linear combination of the system states $x_k^m$ given by a general representation $z_k^m$ such as
\begin{align}\begin{split}\label{eq:state_z}
z_k^m = L_k^m x_k^m.
\end{split}\end{align}
The measure of performance (objective) is given as follows:
\begin{align}\begin{split}
&J=\\
&\cfrac{\Sigma_{k=1}^{K} \Sigma_{m=1}^{M_k} ||z_k^m - \hat{z}_k^m||_{Q_k^m}^2}{||x_0\!-\!\hat{x}_0||_{p_0^{\!\!-\!1}}^2 \!\!+\! \Sigma_{k=1}^{K}\Sigma_{m=1}^{M_k}\left[||w_{k+1}^{m\alpha}||_I^2\!+\!||w_{k+1}^{m\beta}||_I^2\!+\!||v_k^m||_{\!R_k^{-\!m}}^2\!\right]},
\end{split}\end{align}
where none of the denominator terms is exactly zero, $\hat{x}_0$ is an \emph{a priori} estimate of $x_0$, $Q_k^m \geq 0$, $p_0^{-1} > 0$, $I$ identity, and $R_k^m > 0$ are the weighting matrices. The optimal estimate of $z_k^m$, for a given attenuation $\gamma>0$, should satisfy \[\sup J < 1/\gamma.\]

We can rewrite the objective as a minimax problem
\begin{align}
\begin{split}\label{eq:cost}
\min_{\hat{x}_k^m} \max_{y_k^m\!,w_k^m\!,x_0} J = \frac{1}{2} \sum_{k=1}^{K} \sum_{m=1}^{M_k} \{ ||x_k^m - \hat{x}_k^m||_{\bar{Q}_k^m}^2\!\! -\frac{1}{\gamma}(||w_{k+1}^{m\alpha}||_I^2\\+||w_{k+1}^{m\beta}||_I^2+||y_k^m-C_k^m x_k^m||_{R_k^{m^{-1}}}^2)\}  -\frac{1}{2\gamma}||x_0-\hat{x}_0||_{p_0^{-1}}^2
\end{split}
\end{align}
subject to (\ref{eq:model}) where $\bar{Q}_k^m = L_k^{m^T}Q_k^m L_k^m$.

\begin{theorem}
\label{thm}
For noise attenuation $\gamma>0$, an $H_\infty$ filter for $x_k^m$ exists if and only if there exists a stabilizing solution $P_k^m>0~\forall k,m,$ to the following coupled-pair of discrete-domain Riccati equations:
\begin{align}
\begin{split}
\label{eq:riccati}
\!P_{k+1}^{m\alpha} \!=\! A_{k+1}^{m\alpha} P_k^m (I\! -\! \gamma\bar{Q}_k^m P_k^m\!\! +\! C_k^{m^T}\!R_k^{m^{\!\!-1}}\!C_k^m P_k^m )^{-1} \!A_{k+1}^{m\alpha^T}\\+ B_{k+1}^{m\alpha}B_{k+1}^{m\alpha^T},\\
\!P_{k+1}^{m\beta} \!=\! A_{k+1}^{m\beta} P_k^m (I\! -\! \gamma\bar{Q}_k^m P_k^m\!\! +\! C_k^{m^T}\!R_k^{m^{\!\!-1}}\!C_k^m P_k^m )^{-1} \!A_{k+1}^{m\beta^T}\\+ B_{k+1}^{m\beta}B_{k+1}^{m\beta^T},
\end{split}
\end{align}
where $P_0 = p_0$. The $H_\infty$ filter is given by
\[\hat{z}_k^m = L_k^m \hat{x}_k^m,~~~~~k=1,2,\dots,K\]
where \begin{align}\begin{split}\label{eq:mean_upd}
&\hat{x}_{k+1}^{m\alpha*} = A_{k+1}^{m\alpha}\hat{x}_k^{m} + K_{k+1}^{m\alpha} (y_k^{m}-C_k^m \hat{x}_k^{m*}),\\
&\hat{x}_{k+1}^{m\beta*} = A_{k+1}^{m\beta}\hat{x}_k^{m} + K_{k+1}^{m\beta} (y_k^{m}-C_k^m \hat{x}_k^{m*}),\\
\begin{split}
K_{k+1}^{m\alpha} \!=\! A_{k+1}^{m\alpha} P_k^m (I - \gamma\bar{Q}_k^m P_k^m + C_k^{m^T}R_k^{m^{-1}} C_k^m P_k^m )^{-1}\\ C_k^{m^T}R_k^{m^{-1}},
\end{split}\\
\begin{split}
K_{k+1}^{m\beta} \!=\! A_{k+1}^{m\beta} P_k^m (I - \gamma\bar{Q}_k^m P_k^m + C_k^{m^T}R_k^{m^{-1}} C_k^m P_k^m )^{-1}\\C_k^{m^T}R_k^{m^{-1}}.
\end{split}\end{split}
\end{align}
\end{theorem}

{\it Remark 1: }The optimization problem (\ref{eq:cost}) is a zero-sum game with $\hat{x}^m$ as the decision variable of the minimizer player, and $x_0, w_k^m, y_k^m$ as the decision variables of the maximizer player. The game is subject to the dynamic  constraint of (\ref{eq:model}), making it a dynamic game. In the proof for Theorem \ref{thm} (not shown here) we argue that the formulated problem is equivalent to a two-player zero-sum dynamic game, solving which gives us the saddle-point equilibrium solution in (\ref{eq:mean_upd}).

{\it Remark 2:} The proposed estimator is a predictor-based estimator; i.e., it makes use of observations at stage $k-1$ to estimate the states at stage $k$, while not using stage $k$ observations. This approach can degrade the performance substantially because the size of state vectors increases geometrically as the number of stages of the multiscale system increases. Kalman filter implemented in \cite{280746} implicitly uses the current measurements. As an extension here, we now incorporate the observations at the present scale. For linear time-domain systems, Green et. al. \cite{green2012linear} have constructed the $H_\infty$ filter in generalized frameworks. However, for multiscale settings, such extensions are unavailable, and thus we extend the generalized $H_\infty$ filter by Green et. al. \cite[B.3.1]{green2012linear} done for discrete-time systems, to our multiscale state-space setting that also builds upon our main result of Theorem \ref{thm}.

The numerator in our cost function (\ref{eq:cost}) is modified for the estimator based on present scale measurements by making the per-node state-estimation error as follows:
\begin{align*}
\begin{split}
e^m_k = ||\hat{z}^m_{k|k} - L^m_k x^m_k||_{Q_k^m}^2.
\end{split}
\end{align*}
With the new cost function $J$, the optimization problem solution gives the estimator design of (\ref{eq:current_est}). For our experiments, we have used both the predictor-based and current-measurements-based estimators. If measurements at present scale are unavailable, then the results in Theorem \ref{thm} are reduced to:
\begin{align}
\begin{split}
\label{eq:current_est}
\hat{z}_{k|k}^m &= L_k^m\hat{x}_{k|k}^m,\\
\hat{x}_{k|k-1}^{m} &= A_k^{m} \hat{x}_{k-1|k-1}^{m\zeta},\\
P_{k|k-1}^m &= A_{k-1}^{m\zeta} P_{k-1|k-1}^{m\zeta} A_{k-1}^{m\zeta \top} + B_{k-1}^{m\zeta}B_{k-1}^{m\zeta \top},\\
\hat{x}^m_{k|k} &= \hat{x}^m_{k|k-1} + K_k^m (y_k^m - C^m_k x^m_{k|k-1}),\\
K_k^m &= P^m_{k|k-1} C_k^{m\top} (R^m_k + C^m_kP^m_{k|k-1}C_k^{m\top})^{-1},\\
P^m_{k|k} &= P^m_{k|k-1} \!- P^m_{k|k-1} \!\begin{bmatrix}C_k^{m\top}\!\!&\!\!L_k^{m\top}\end{bmatrix}\![\tilde{R}_{k}^{m}]^{-1}\!\begin{bmatrix}C_k^m\\L_k^m\end{bmatrix}\!P^m_{k|k-1},\\
\tilde{R}_{k}^m &= \begin{bmatrix}R_k^m&0\\0&-I/\gamma\end{bmatrix} + \begin{bmatrix}C^m_k\\L^m_k\end{bmatrix}P^m_{k|k-1} \begin{bmatrix}C_k^{m\top}&L_k^{m\top}\end{bmatrix}.
\end{split}
\end{align}

\section{Numerical Experiments}
\begin{figure}[t]
	\centering
	\subfloat{\includegraphics[width=1.25\linewidth]{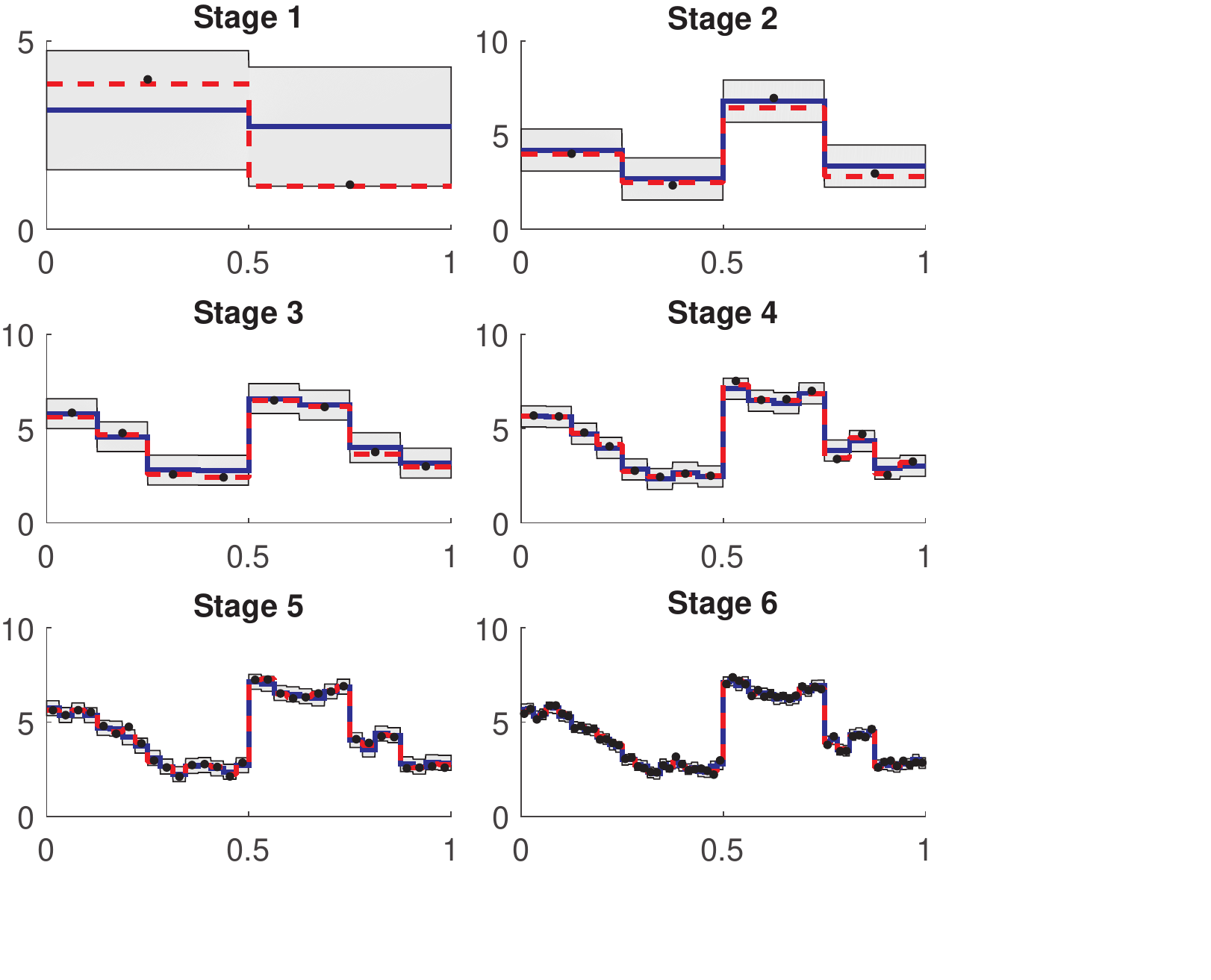}}
\vspace*{-12mm}
\caption{Example of a one-dimensional multiresolution signal represented at six different levels. Gray $70\%$ confidence bounds indicate the uncertainty in state transitions. Observations are indicated by black dots, while the original signal is in blue. State estimates recovered by proposed filter ($\gamma=1$) from observations at all resolution levels are indicated by red dotted line.\label{fig:1D_signal}}
\end{figure}

\begin{figure}[t]
	\centering
	\includegraphics[width=1\linewidth]{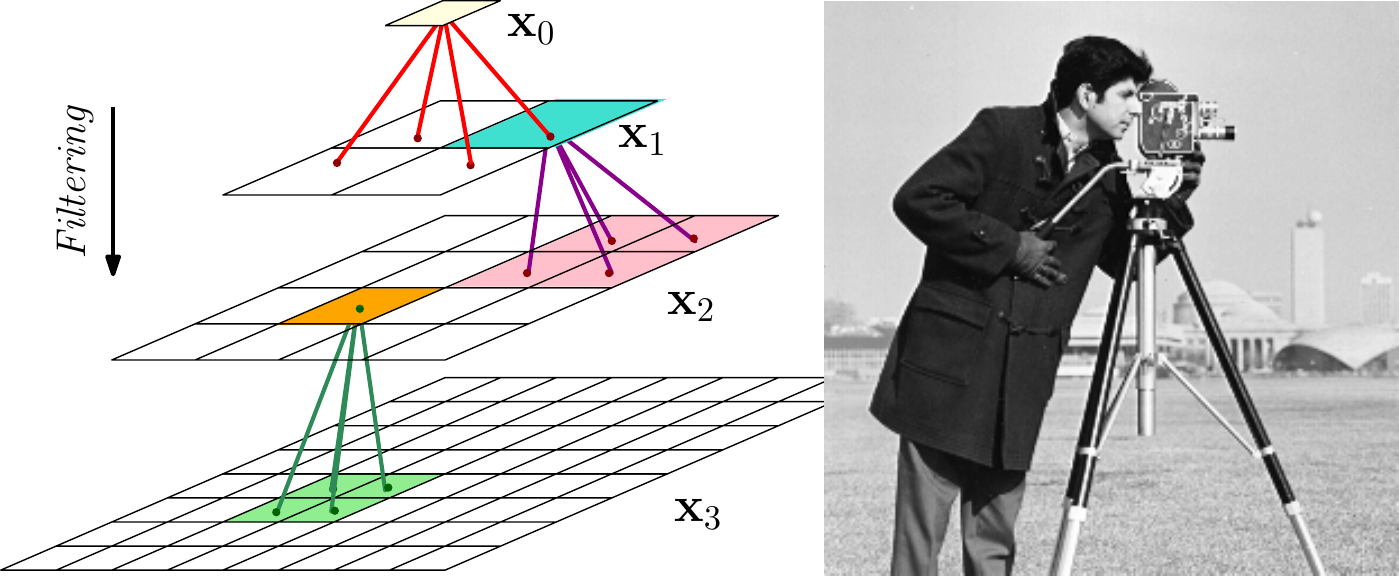}
\caption{Hierarchical multiscale structure for 2-D signals (left) and the cameraman image that we take as the finest resolution signal of our multiscale signal to use in experiments (right).}
\label{fig:2d_model}
\end{figure}
\begin{figure}[t]
	\centering
	\includegraphics[width=1\linewidth]{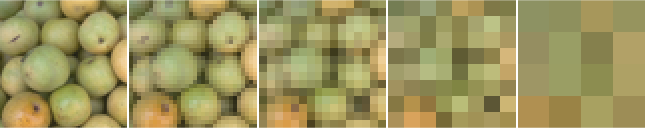}
\caption{Different resolution levels of a 2-D signal constructed by recursive averaging as shown in Fig.~\ref{fig:2d_model}. These images after adding noise will serve as the observations for the state-estimator to recover original image.}\label{fig:2d_signal}
\end{figure}
\begin{figure}[t]
	\centering
	\includegraphics[width=1\linewidth]{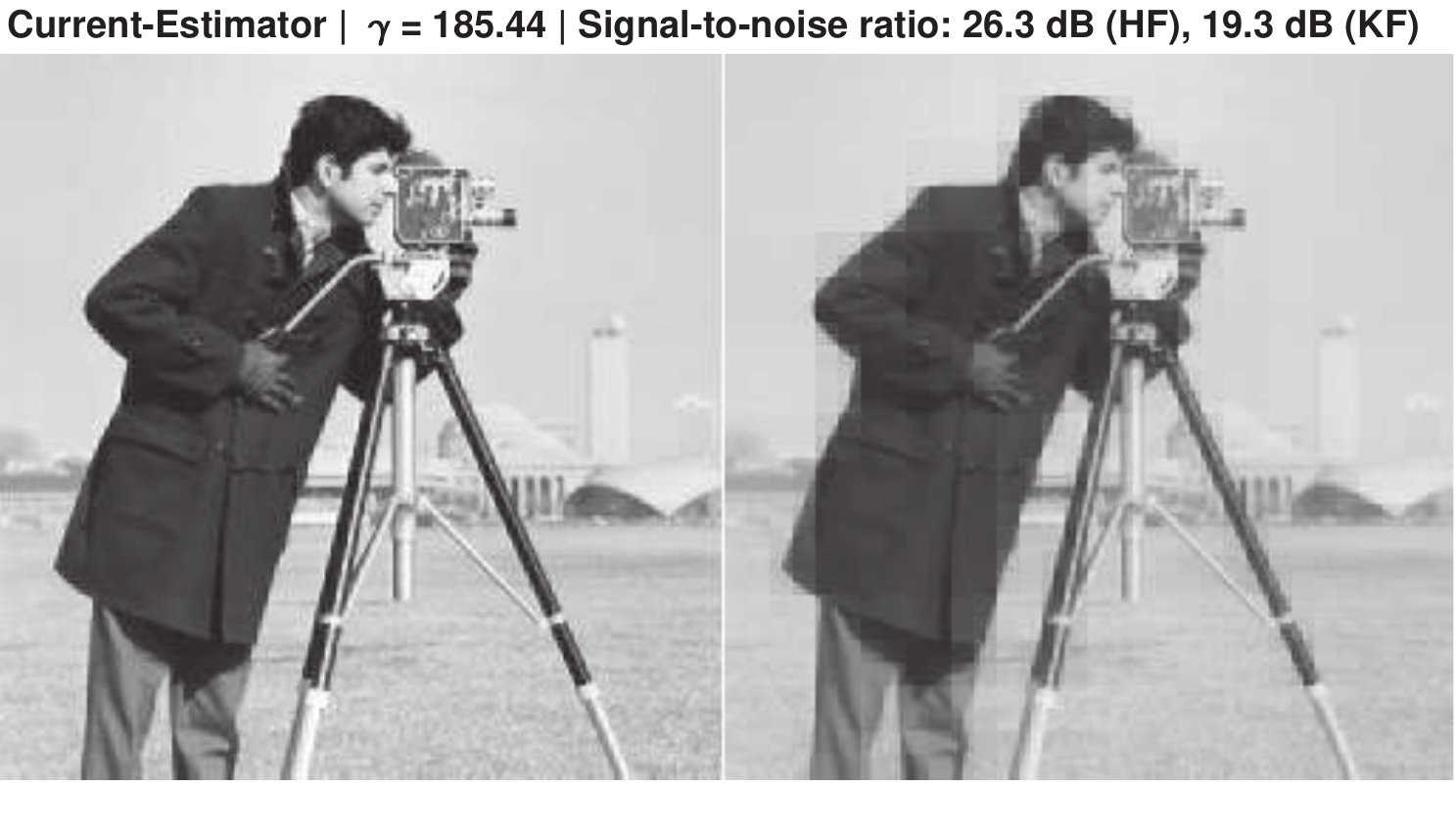}
	\vspace*{-8mm}
\caption{Estimated image at finest resolution level using $H_\infty$ filter with $\gamma = 185.44$ (left) and Kalman filter (right). High SNR for $H_\infty$ estimate corresponds to a better image as is shown: block-like artifacts are visible near the edges in the image on right.}
\label{fig:HFvsKFresult}
\end{figure}
Fig.~\ref{fig:1D_signal} shows various levels of a multiscale process on the interval $[0,1]$. The process is a step function, and each sub-interval corresponds to a node of an autoregressive process on a dyadic tree. We show that as more stages are observed, the confidence interval around state estimates (red) becomes tighter giving more precision and they also tend to coincide with original signal (blue) showing accuracy. We let $A$, $B$ and $C$ be identity matrices in all our experiments for simplicity.

2-D signals have similar hierarchical modeling structure compared to 1-D signals, see Fig.~\ref{fig:2d_model}. The computation overload is also not substantial because the way we model it requires more children per node. We test our algorithm on different RGB and grayscale images. The first step we take into account before estimation is the construction step that takes the finest image, i.e., the actual image, and generates its low-resolution versions step by step to yield images at all tree levels, Fig.~\ref{fig:2d_signal}. Unlike our 1-D examples, here, we assume that the original signal exists in the continuous real domain, i.e., as $K\to \infty$, $x_K$ approaches the true signal.

A series of experiments have been performed with 2-D image data. To generate a layered data of this signal at lower resolution scales, we have averaged over a neighborhood of four pixels each recursively in a pyramid-like structure, see Fig.~\ref{fig:2d_model}. Other methods such as frequency-domain representations can also be used, but they will not affect our results. Once we have the data at various levels, we use it to perform filtering by giving different noisy and corrupted observations to our system. In Fig.~\ref{fig:HFvsKFresult}, we show that our method outperforms the standard Kalman filtering approach for multiscale models, giving higher SNR values and less block-like artifacts. Here, the cameraman image at $256\times 256$ resolution serves as the finest resolution signal. 

\section{Discussion and Conclusion}
\begin{figure}[t]
	\centering
\includegraphics[width=0.85\linewidth]{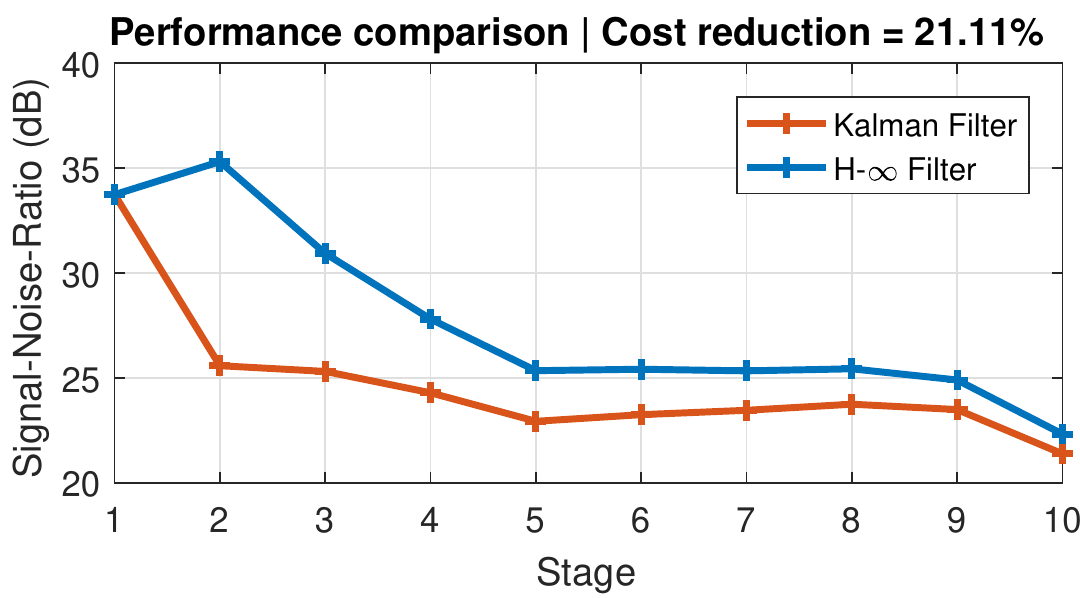}
\caption{Comparison of performance in cost reduction for standard Kalman and the proposed estimators ($\gamma = 84.44$).\label{fig:snrtrend}
\vspace*{-2mm}}
\end{figure}
\begin{figure}[t]
	\centering
	\subfloat{\includegraphics[width=0.5\linewidth]{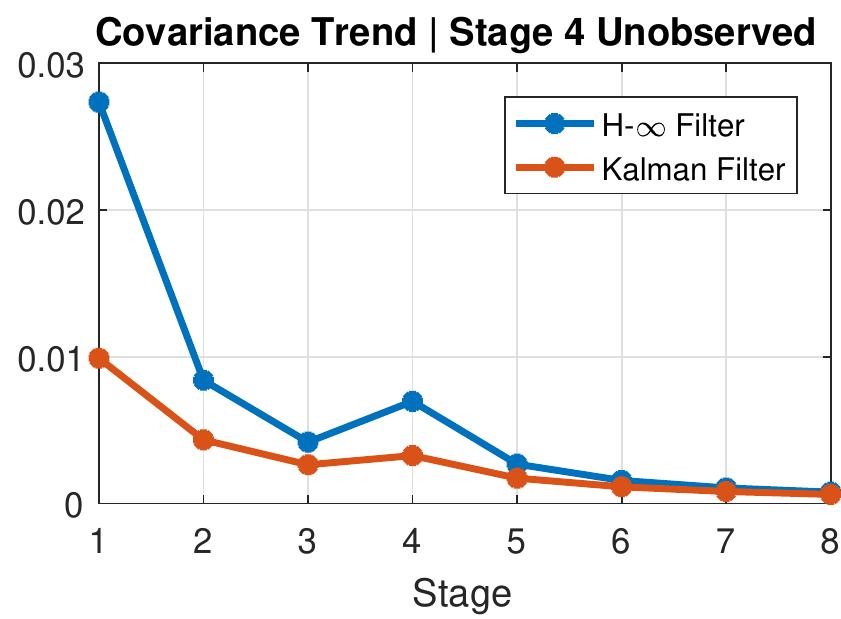}}
\hfill
\subfloat{\includegraphics[width=0.5\linewidth]{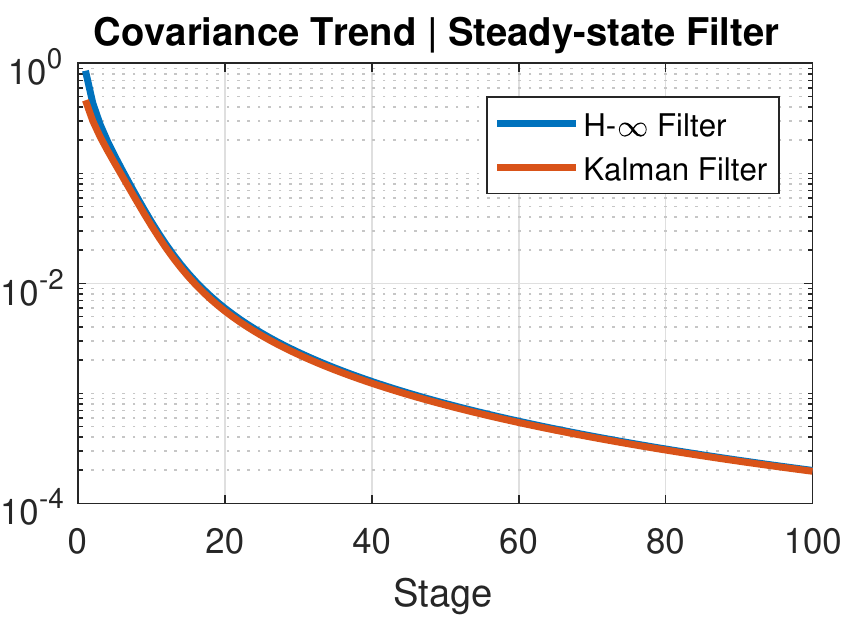}\vspace*{-5mm}}
\vspace*{-2mm}
\caption{Trend of uncertainty (covariance) in state-estimates is shown. When there is no observation recorded at a given stage, uncertainty in the state estimate of that stage increases (left). As the system reaches the steady-state, the covariance is shown to decrease (right).\label{fig:missingObs_ss}}
\end{figure}
Using the minimax game-theoretic approach to the optimal estimation problem, we have avoided solving the smoothing and filtering problems separately that has classically been considered in multiscale recursive estimation \cite{280746,nagpal1991filtering}. For our estimator, we have avoided the reverse (fine-to-coarse) dynamics by only considering natural dynamics. Methodology employed by Chou et. al. \cite{280746} uses reverse dynamics as it involves two-sweep estimation process: first filtering and then smoothing. The reason for doing so is to allow the measurement at one node to contribute towards the estimate at another node on the same resolution level. In our approach, this technicality is taken care of by the reverse Riccati recursion in (\ref{eq:riccati}).

In Fig.~\ref{fig:snrtrend}, we show that by using $H_{\infty}$ filter instead of standard Kalman filter, a large reduction in the estimation cost of $21\%$ is achieved (i.e. calculated by accumulating squared differences between original and state estimates). Across all the stages, we see that $H_{\infty}$ filter exhibits a higher value of signal-to-noise ratio in comparison with Kalman filter.

One important feature of the multiscale filter  is the scenario of missing information at different nodes across the tree. The information at neighboring nodes can improve the state estimate of nodes where no information is recorded. We perform an experiment in which observations have not been recorded for any node of stage $4$. The estimators can still give considerably good state estimates for nodes at stage $4$, with a yet higher uncertainty. This increase in estimation uncertainty is shown in Fig.~\ref{fig:missingObs_ss}. Although we have observed a lower variance for Kalman filter,  the variance converges in Fig.~\ref{fig:missingObs_ss} as we move towards finer stages. Note that when the proposed $H_\infty$ filter is used, the covariance update in (\ref{eq:riccati}) is not the mere Lyapunov update because of a nonzero $\gamma$. However, the mean update in (\ref{eq:mean_upd}) is predicted based on state matrix $A$ only.

In this paper, we have presented an optimal robust filter design for a class of multiscale discrete domain systems that combats worst-case additive exogenous process and sensor noise. Our filter design gives the saddle-point solution which is the equilibrium strategy of the minimax two-player zero-sum dynamic game. We have given the conditions for an optimal solution both for the predictor-based estimator (\ref{eq:mean_upd}) and the current-measurements-based estimator (\ref{eq:current_est}) for a given disturbance attenuation level $\gamma$. Experimental results corroborate our theoretical findings, and we have shown our evaluations for 1-D and 2-D signal examples. Our work is not limited to the scenarios of images and one-dimensional signals, and in fact, we have argued that for the multiscale state-space modeling framework, the application of this work is not limited by computational or memory constraints. Moreover, the expressiveness of this model is used in various domains such as graphical topic modeling in machine learning \cite{ahmed2013hierarchical}. Multi-resolution phenomena are widely found in numerous IoT-related applications. And this fact makes our work useful for different domains. We also show the convergence of state trajectories in steady-state and a comparative analysis with the standard Kalman filter. Lastly, dealing with missing information and the distributed nature of such models would be a future direction of our work.

\bibliographystyle{IEEEbib}
\bibliography{strings,refs}

\end{document}